\documentclass{iopart}
\usepackage{iopams}
\usepackage{bm,mathrsfs}
\usepackage[mathcal]{euscript}
\usepackage{temp}
\usepackage[breaklinks=true,unicode=true,urlcolor = blue,colorlinks = true,citecolor = blue,linkcolor = blue]{hyperref}
\usepackage{pgf,tikz}
\usepackage{graphicx}
\usepackage{subcaption}
\usepackage[numbers,sort&compress]{natbib}
\renewcommand{\vec}[1]{\bm{#1}}
\renewcommand{\imath}{\rmi}
\begin{document}
\title{Curvature effects in statics and dynamics of low dimensional magnets}
\author{Denis D. Sheka$^1$, Volodymyr P. Kravchuk$^2$ and Yuri Gaididei$^2$}
\address{$^1$ Taras Shevchenko National University of Kiev, 01601 Kiev, Ukraine}
\address{$^2$ Bogolyubov Institute for Theoretical Physics, 03143 Kiev, Ukraine}

\eads{\mailto{sheka@univ.net.ua}, \mailto{vkravchuk@bitp.kiev.ua}, \mailto{ybg@bitp.kiev.ua}}

%
%%%%%%%%%%%%%%%%%%%%%%%%%%%%%%%%%%%%%%%%%%%%%%%%%%%%%%%%%%%%%%%%%%%%%70
%
%         ABSTRACT
%
%%%%%%%%%%%%%%%%%%%%%%%%%%%%%%%%%%%%%%%%%%%%%%%%%%%%%%%%%%%%%%%%%%%%%70
%
\begin{abstract}
We develop an approach to treat magnetic energy of a ferromagnet for arbitrary curved wires and shells on the assumption that the anisotropy contribution much exceeds the dipolar and other weak interactions. We show that the curvature induces two effective magnetic interactions: effective magnetic anisotropy and effective Dzyaloshinskii-like interaction. We derive an equation of magnetisation dynamics and propose a general static solution for the limit case of strong anisotropy.  To illustrate our approach we consider the magnetisation structure in a ring wire and a cone surface: ground states in both systems essentially depend on the curvature excluding strictly tangential solutions even in the case of strong anisotropy. We derive also the spectrum of spin waves in such systems.
\end{abstract}
\pacs{75.30.Et, 75.75.-c, 75.78.-n}
%\keywords{classical spin models,curvature,nanowire, nanoshell}
%\submitto{\jpa}
% 75.30.Et	Exchange and superexchange interactions (see also 71.70.Gm Exchange interactions)
% 75.75.-c	Magnetic properties of nanostructures
% 75.78.-n	Magnetization dynamics
%
\date{January 27, 2015}
\maketitle
\section{Introduction}
\label{sec:intro}

An interplay between topology of the order parameter field and the geometry of the underlying substrate attracts attention of many researchers in the modern physics of condense matter and in field theories. One of the well--known examples of the nonlinear vector field model is a general Ginzburg--Landau vector model with the energy functional \cite{Pismen99}
\begin{equation}\label{eq:GL}
E = \int \mathrm{d}\vec{x} \left[  \vec{\nabla}\vec{u} : \vec{\nabla}\vec{u} + V(\vec{u})\right]
\end{equation}
for the vector order parameter $\vec{u}=\left(u^1,u^2,\dots,u^n\right)$ in the multidimensional real space $\vec{x}\in\mathrm{R}^d$, where double dot denotes a scalar product in both real and order parameter spaces. If transformations of the order parameter and transformations of the real space are independent the double dot scalar product computed in such a way that the vector components in both spaces do not mix: $ \vec{\nabla}\vec{u} : \vec{\nabla}\vec{u}=G_{ij}\,\nabla u^i\,\nabla u^j $  with $G_{ij}$ being a metric tensor of the order parameter space. A behaviour of vector fields in a curved space \cite{Bowick09,Turner10} becomes more sophisticated due to the intimate relation between the geometry of the substrate space ($\vec{x}$--variable) and the geometry of the field ($\vec{u}$--variable). In most studies the vector field was supposed to be strictly tangential to the curved substrate. For example, this assumption was used when the role of curvature  in the interaction between defects was studied in 2D $XY$-like models, which can describe thin layers of superfluids, superconductors, and liquid crystals deposited on curved surfaces  \cite{Vitelli04}.

Nowadays there is a growing interest to low dimensional magnetic objects such as magnetic nanoshells and nanowires. This interest is supported by a great advantage in nanotechnology, including numerous magnetic devices (high-density data storage, logic, sensing devices etc). Theoretical description of the evolution of magnetisation structures in such systems is based on the dynamics of three dimensional (3D) vector order parameter such as magnetisation unit vector $\vec{m}$ in a constrained physical space, e.g. quasi two dimensional (2D) nanoshell and quasi one dimensional (1D) nanowire. The interrelation between the topological properties of magnetic structures and the underlying curvature complicates an analysis, nevertheless it can be a source of new effects. For example, $2\pi $--skyrmions can appear in Heisenberg isotropic magnets due to a coupling between magnetic field and curvature of the surface \cite{Carvalho-Santos12}; in easy-surface Heisenberg magnets the curvature of the underlying surface leads to a coupling between the localised out-of-surface component of magnetic vortex with its delocalised in-surface structure \cite{Kravchuk12a}. It is well known that the curvature of the system can induce an additional effective energy contribution, a so-called `geometrical potential': In  a seminal paper \cite{Costa81} da Costa  developed a quantum mechanical approach to study the tangential motion of a particle rigidly bounded to a surface. Similar `geometrical potential' appears in 1D curved quantum wires \cite{Magarill05}. Effects of effective anisotropy induced by the curvature were also discussed for quasi 1D curved magnetic nanowires of particular geometries \cite{Tkachenko12,Tkachenko13}. In general, the influence of the `geometrical potential' is ``proportional to the second degree of curvature of the system'' \cite{Magarill05}. In spite of numerous results on the behaviour of vector field in curved systems, see e.g. review articles \cite{Bowick09,Turner10}, the problem is not fully understood. In particular, the 3D vector field in majority of studies was assumed to be rigidly bound to the surface in the case of magnetic shells or to the curve in the case of magnetic wires.

Very recently we have developed fully 3D approach for thin magnetic shells of arbitrary shape \cite{Gaididei14}. In this paper we extend this approach for both 2D shells and 1D wires. We base our study on the phenomenological Landau-Lifshitz equation
\begin{equation} \label{eq:LL}
\partial_t \vec m = \omega _0\, \vec m\times\frac{\delta E}{\delta \vec m},
\end{equation}
which describes the classical magnetisation dynamics. Here $E$ is the total energy normalised by $4\pi M_s^2$ with $M_s$ being the saturation magnetisation, and characteristic time scale of the system is determined by the frequency $\omega _0=4\pi \gamma_0 M_s$ with $\gamma_0$ being the gyromagnetic ratio. The damping is neglected. For an arbitrary orthogonal basis $\{\vec e_1,\,\vec e_2,\,\vec e_3\}$ one can parameterise the unit magnetisation vector as follows
\begin{equation} \label{eq:theta-phi-1D}
\vec{m} = \sin\theta \cos\phi \, {\vec{e}}_1 + \sin\theta \sin\phi \,{\vec{e}}_2 + \cos \theta \, {\vec{e}}_3,
\end{equation}
where angular variables $\theta $ and $\phi $ depend on spacial and temporal coordinates. Within the angular representation \eqref{eq:theta-phi-1D} the equation of motion \eqref{eq:LL} reads
\begin{equation} \label{eq:LL-angular}
\sin\theta  \partial_t \phi =\omega _0\,\frac{\delta E}{\delta \theta }, \qquad - \sin\theta  \partial_t \theta  = \omega _0\,\frac{\delta E}{\delta \phi }.
\end{equation}
The total energy of the magnet can collect different contributions such as energies of exchange, anisotropy and dipolar one. In the following we consider a hard magnet where the anisotropy contribution much exceeds the dipolar and other weak interactions.
Therefore in the current study we restrict ourselves to the consideration of Heisenberg magnets. In this case the total energy functional has the following form:
\begin{equation} \label{eq:E-total}
E = \int \mathrm{d}V \left[ \ell^2\mathscr{E}_{\mathrm{ex}} + \lambda  \left(\vec{m}\cdot\vec{n}\right)^2\right].
\end{equation}
Here the first term describes the isotropic exchange interaction, see below Eq. \eqref{eq:Eex-Cartesian}, with $\ell=\sqrt{A/(4\pi  M_s^2)}$ being an exchange length and $A$ being an exchange constant. The second term in \eqref{eq:E-total} is the energy of anisotropy. The unit vector $\vec{n}$ gives the direction of the anisotropy axis and $\lambda $ is a dimensionless anisotropy constant.

An important point is that the vector $\vec n(\vec r)$ is a function of spatial coordinates in accordance with the geometry of the curvilinear sample. For example, if $\vec n$ is normal to a curvilinear shell and $\lambda >0$ then we have a case of easy-surface anisotropy; if $\vec n$ is tangential to a curvilinear wire and $\lambda <0$, then we have a case of easy-tangential anisotropy, etc. Actually, this is the way how the curvature is introduced to the problem.

The anisotropic curvilinear systems \eqref{eq:E-total} with a nontrivial topology are of particular interest, since topologically nontrivial magnetisation distributions are inherent here. Examples are magnetic vortices in easy-surface spherical shells \cite{Kravchuk12a}, magnetic domains in M\"obius rings with easy-normal anisotropy \cite{Yoneya08,Pylypovskyi14d}.

A natural way to look for magnetisation distributions in the curvilinear systems is to proceed to the corresponding curvilinear basis. However the representation of the exchange contribution $\mathscr{E}_{\mathrm{ex}}$ in an \emph{arbitrary} curvilinear frame of reference is a quite challenge. Previously this problem was solved for a couple of simple geometries, namely, cylindrical \cite{Landeros10} and spherical \cite{Kravchuk12a}. In the present work we propose a general approach to derive the exchange energy for arbitrary curvilinear 1D and 2D systems (curved wires and curved shells) and an arbitrary magnetisation vector field, not necessarily tangential to the surface as it was recently considered for an nematic shells \cite{Napoli12,Napoli12a}. We show that in the curvilinear systems there appear two effective magnetic interactions: (i) curvature induced effective anisotropy which is bilinear with respect to the curvature and the torsion and is similar to the `geometrical potential', (ii) curvature induced effective Dzyaloshinskii-like interaction, which is linear with respect to the curvature and the torsion.

The paper is organised as follows. We derive the energy of the curved 1D wire in Section \ref{sec:1D}; our approach is illustrated by the calculation of the ground state of a highly anisotropic curved wire. In Section \ref{sec:2D} we discuss the role of the effective anisotropy and the effective Dzyaloshinskii-like interaction for 2D curved shell. We consider two applications  of our theory: the ground state of the narrow ring wire and the spectrum of spin-waves are calculated in Section \ref{sec:ring}, the spin-wave spectrum for the cone shell is derived in Section \ref{sec:cone}. In Section \ref{sec:discussion} we present some remarks about possible perspectives. Another representation of the energy of 1D wire is proposed in \ref{sec:energy-1d-another}.

\section{Energy and curvature induced effective fields for a curved wire}
\label{sec:1D}

We start with a one-dimensional case and consider a thin nanowire whose transverse size is small enough to ensure the magnetisation uniformity along the crosswise direction. One can describe a wire using Frenet–-Serret parametrisation for a 3D curve $\vec{\gamma }$. We use its natural parametrisation by arc length $s$ of general form $\vec{\gamma }  = \vec{\gamma }(s)$. In Cartesian basis $\hat{\vec x}_i\in\{\hat{\vec x},\,\hat{\vec y},\,\hat{\vec z}\}$, one can parameterise the curve as $\vec{\gamma }  = \gamma _i \hat{\vec{x}}_i$. The Einstein summation convention is used here and everywhere below. Let us introduce the local normalised curvilinear basis (Frenet–-Serret frame):
\begin{equation} \label{eq:1D-basis}
\vec{e}_1 = \vec{\gamma }', \qquad \vec{e}_2 = \frac{\vec{e}_1'}{\left|\vec{e}_1' \right|}, \qquad \vec{e}_3 = \vec{e}_1\times \vec{e}_2
\end{equation}
with $\vec{e}_1$ being the tangent, $\vec{e}_2$ being the normal, and $\vec{e}_3$ being the binormal to the curve $\vec{\gamma }$. Here and below the prime denotes the derivative with respect to the arc length $s$. Note that $|\vec{\gamma }'(s)|=1$ in the natural parametrisation. The differential properties of the curve are determined by Frenet–-Serret formulae:
\begin{equation} \label{eq:Frenet–Serret}
\vec{e}_\alpha ' = F_{\alpha \beta }\vec{e}_\beta , \qquad \left\|F_{\alpha \beta } \right\| =
\left(
\begin{array}{ccc}
0 & \kappa & 0 \\
-\kappa & 0 & \tau  \\
0 &- \tau  & 0
\end{array}
\right),
\end{equation}
where $\kappa $ is the curvature of the wire and $\tau $ is its torsion. Latin indices $i,j=1,2,3$ describe the Cartesian coordinates and the Cartesian components of vector fields, whereas Greek indices $\alpha ,\beta =1,2,3$ numerate the curvilinear coordinates and the curvilinear components of vector fields.

After defining the 3D curve one can parameterise a physical wire with a finite crosswise size. We take $\vec{\gamma }(s)$ as the central curve of a wire. Then the space domain filled by the wire can be parameterised as
\begin{equation} \label{eq:3D-curve}
\vec{r}(s,\xi _2,\xi _3) = \vec{\gamma }(s) + \xi _2 \vec{e}_2 + \xi _3 \vec{e}_3,
\end{equation}
where $\vec{\xi } = (\xi _2,\xi _3)$ are coordinates within the cross section, $|\vec{\xi }|\lesssim h$ with $h$ being the wire thickness. The assumption of the magnetisation one-dimensionality can be formalised as $\vec m=\vec m(s)$. This assumption is appropriate for the cases when the thickness $h$ does not exceed the characteristic magnetic length. We also suppose that $h\ll1/\kappa ,1/\tau $.

We base our study on the the energy functional \eqref{eq:E-total}. In the Cartesian frame of reference the exchange energy density has the form
\begin{equation} \label{eq:Eex-Cartesian}
\mathscr{E}_{\mathrm{ex}} = \left( \vec{\nabla} m_i \right) \left( \vec{\nabla} m_i \right).
\end{equation}
Now one can express the Cartesian components of the magnetisation vector $m_i$  in terms of the curvilinear components $m_\alpha$ as follows
\begin{equation} \label{eq:m-Cartesian}
m_i=m_\alpha  \left({\vec e}_\alpha \cdot\hat{\vec x}_i\right).
\end{equation} Then we substitute this expression into $\mathscr{E}_{ex}$ and apply del operator in its curvilinear form, $\vec\nabla\equiv {\vec e}_1 \partial_s$. Finally the energy density in the Frenet–-Serret frame of reference reads
\begin{subequations} \label{eq:E-1D}
\begin{equation} \label{eq:E-sum-1D}
\mathscr{E}_{\mathrm{ex}} = \left( m_\alpha  \vec{e}_\alpha \right)'\left( m_\beta  \vec{e}_\beta \right)' = \mathscr{E}_{\mathrm{ex}}^{0}  + \mathscr{E}_{\mathrm{ex}}^{A} + \mathscr{E}_{\mathrm{ex}}^{D}.
\end{equation}
Here the first term describes the common isotropic part of exchange expression which has formally the same form as for the straight wire
\begin{equation} \label{eq:E-1D-0}
\mathscr{E}_{\mathrm{ex}}^{0} = m_\alpha 'm_\alpha ' = |\vec{m}'|^2.
\end{equation}
The second term $\mathscr{E}_{\mathrm{ex}}^{A}$ describes an effective anisotropy--like interaction,
\begin{equation} \label{eq:E-1D-A}
\fl
\mathscr{E}_{\mathrm{ex}}^{A} = K_{\alpha \beta } m_\alpha  m_\beta , \quad K_{\alpha \beta } = F_{\alpha \gamma }F_{\beta \gamma },\quad \left\|K_{\alpha \beta }\right\| = \left(
\begin{array}{ccc}
\kappa ^2  & 0       & -\kappa \tau  \\
0    & \kappa ^2+\tau ^2 & 0  \\
-\kappa \tau   & 0       & \tau ^2
\end{array}
\right).
\end{equation}
Components of the tensor $K_{\alpha \beta }$ are bilinear with respect to the curvature $\kappa $ and the torsion $\tau $, they play the role of effective anisotropy coefficients, induced by the curvature of the wire. In some sense it is similar to the `geometrical potential' which an electron experiences  in a curved quantum wire \cite{Magarill05}. Note that an effective `geometrical' magnetic field was calculated recently in curved magnonic waveguides \cite{Tkachenko12}. As opposed to previous studies we will show below that the curvature and torsion can cause a new magnetisation ground state, which is absent in the straight case.

The last term \eqref{eq:E-1D-A} is a combination of Lifshitz invariants
\begin{equation} \label{eq:E-1D-D}
\mathscr{E}_{\mathrm{ex}}^{D} = F_{\alpha \beta }\left(m_\alpha  m_\beta '-m_\alpha ' m_\beta \right)
\end{equation}
\end{subequations}
and therefore it can be interpreted as an effective Dzyaloshinskii interaction
\cite{Dzyaloshinsky57,Dzyaloshinsky58}. The tensor of coefficients of this effective Dzyaloshinskii interaction $F_{\alpha \beta }$ is exactly the Frenet transformation matrix, see \eqref{eq:Frenet–Serret}, which is linear with respect to the curvature $\kappa $ and the torsion $\tau $. Due to the linear form this term can cause phenomena, which depend on the sign of $\kappa $ and $\tau $. %In particular, we will see below that namely due to $\mathscr{E}_{\mathrm{ex}}^{D}$ there appears the linear term in the spin wave spectrum for a helix nanowire, leading to the zero group velocity at finite wave vector.

Using the angular notations \eqref{eq:theta-phi-1D} one can rewrite the energy terms as follows:
\begin{equation} \label{eq:E-1D-theta-phi}
\eqalign{
\mathscr{E}_{\mathrm{ex}}^{0} &= \theta'^2 + \sin^2\theta \phi'^2,\\
\mathscr{E}_{\mathrm{ex}}^{A} &=(\kappa \sin\theta -\tau \cos\theta \cos\phi )^2+\tau ^2\sin^2\phi \\
\mathscr{E}_{\mathrm{ex}}^{D} &=\phi '\left(2\kappa \sin^2\theta -\tau \sin2\theta \cos\phi \right)-2\tau \theta '\sin\phi .
}
\end{equation}
Finally, by summing up all terms in \eqref{eq:E-1D-theta-phi}, we get
\begin{equation} \label{eq:E-1D-final}
\mathscr{E}_{\mathrm{ex}}^{1d} = \left[\theta '\!-\tau \sin\phi \right]^2\!\!\! + \left[\sin\theta  (\phi '+\kappa )-  \tau \cos\theta \cos\phi  \right]^2\!\!.
\end{equation}
Note that the exchange magnetic energy of curved wire was recently calculated in Ref.~\cite{Slastikov12}. However curvature effects were ignored and the exchange energy was written in the form $\mathscr{E}_{\mathrm{ex}}^{0}$. For some applications, e.g. studying domain walls dynamics, it is useful to rewrite the energy \eqref{eq:E-1D-final}, using another angular parametrisation, where the polar angle $\theta $ is counted from the tangential direction, see \ref{sec:energy-1d-another}.

Let us take into account the anisotropy term in \eqref{eq:E-total}. We choose the anisotropy axis along the central line of the wire, $\vec{n}=\vec{e}_1$. The total energy density, according to \eqref{eq:E-total} has the form:
\begin{equation} \label{eq:Eex+Ea-1D}
\mathscr{E}^{1d} =  \ell^2\mathscr{E}_{\mathrm{ex}}^{1d}+ \lambda \sin^2\theta \cos^2\phi .
\end{equation}

The developed approach enables us to obtain a general static solution for the high-anisotropy case. We consider a physically interesting case of easy-tangential anisotropy $\lambda <0$, which favours the magnetisation distribution tangential to the wire. In the strong anisotropy limit the magnetisation is quasitangential, therefore $\theta =\pi /2+\vartheta $, and $|\vartheta |, |\phi |\ll1$. Then the total energy density can be rewritten as follows
\begin{equation} \label{eq:En-1D+anis}
\mathscr{E}^{1d} \approx \mathscr{E}_{\mathrm{ex}}^{0} + \underbrace{2\ell^2(\tau \kappa \vartheta -\kappa '\phi )}_{\mathscr{E}^F}+|\lambda |(\vartheta ^2+\phi ^2)+ \mathrm{const},
\end{equation}
where the second summand is the energy density of strictly tangential distribution, the third summand can be written as $\mathscr{E}^F=-(\vec F\cdot\vec m)$ and therefore one can consider it as an interaction with an effective curvature induced magnetic field
\begin{equation} \label{eq:field-1D}
\vec F=2\ell^2 \left(\kappa '\vec e_2+\tau \kappa \vec e_3\right),
\end{equation}
and the last summand in \eqref{eq:En-1D+anis} represents the anisotropy contribution. Minimisation of the energy functional \eqref{eq:En-1D+anis} with respect to $\vartheta $ and $\phi $ results in
\begin{equation} \label{eq:gs-1D}
\theta  = \frac{\pi }{2} - \frac{\ell^2}{|\lambda |} \kappa \tau  + \mathcal{O}\left(\frac{1}{|\lambda |^2}\right), \qquad \phi  = \frac{\ell^2}{|\lambda |} \kappa ' + \mathcal{O}\left(\frac{1}{|\lambda |^2}\right).
\end{equation}
According to \eqref{eq:gs-1D} the strictly tangential solution is realised only for a specific case $\tau =0$ and $\kappa '=0$. Note that in the main part of recent studies of magnetisation states in curved nanowires the tangential magnetisation distributions were considered only \cite{Tkachenko12,Tkachenko13}. The solution for the ground state of 1D magnets \eqref{eq:gs-1D} is in agreement with recent results for 2D surfaces \cite{Gaididei14}.

\section{Energy and the curvature induced effective fields for a curved shell}
\label{sec:2D}

In this section we consider the curvature induced effects in magnetic nanoshell using a thin--shell limit. We describe a shell considering a surface $\vec{\varsigma }(\xi _1,\xi _2)$ with $\xi _1$ and $\xi _2$ being local curvilinear coordinates on the surface. In the Cartesian basis $\hat{\vec x}_i\in\{\hat{\vec x},\,\hat{\vec y},\,\hat{\vec z}\}$, one can parameterise the surface as $\vec{\varsigma }  = \sigma _i \hat{\vec{x}}_i$. We define the local curvilinear basis as follows:
\begin{equation} \label{eq:2D-basis}
\vec{e}_1 = \frac{\vec{g}_1}{|\vec{g}_1|}, \quad \vec{e}_2 = \frac{\vec{g}_2}{|\vec{g}_2|}, \quad \vec{e}_3 = \vec{e}_1\times \vec{e}_2, \quad \vec{g}_\mu  = \partial_\mu  \vec{\varsigma },
\end{equation}
where $\partial_\mu =\partial/\partial\xi_\mu $ with $\mu =1,2$. Similar to notations of the previous section, Greek indices $\alpha ,\beta ,\gamma =1,2,3$ numerate curvilinear coordinates and curvilinear components of vector fields. To indicate only in--surface curvilinear coordinates we use Greek notations $\eta , \mu , \nu  = 1,2$. We suppose that the surface curvilinear frame $\{\vec{e}_1,\vec{e}_2\}$ is orthogonal one, hence the surface metric tensor $g_{\mu \nu } = \vec{g}_\mu \cdot \vec{g}_\nu $ has a diagonal form, $\left\|g_{\mu \nu }\right\| = \mathrm{diag}\left(g_{11},g_{22}\right)$. The local curvilinearity of the basis $\vec{e}_\mu $ is determined by the second fundamental form $b_{\mu \nu } = \vec{e}_3\cdot \partial_\nu \vec{g}_\mu $. The Gauss curvature is $\mathcal{K}=\det(H_{\alpha \beta })$ and the mean curvature is $\mathcal{H}=\mathrm{Tr}(H_{\alpha \beta })/2$ with the Hessian matrix given by $||H_{\mu \nu }|| = ||b_{\mu \nu }/\sqrt{g_{\mu \mu }g_{\nu \nu }}||$.

The differential properties of the curvilinear basis are determined by Gauss--Codazzi equation
\begin{equation} \label{eq:Gauss}
\partial_\mu \vec{g}_\nu = b_{\mu \nu } \vec{e}_3 + \left\{\eta \atop\mu \nu \right\} \vec{g}_{\eta }
\end{equation}
with $\left\{\eta \atop \mu \nu \right\}$ being the Christoffel symbol.

Let us parameterise the ferromagnetic shell using the thin--shell limit. Considering the surface $\vec{\varsigma }(\xi _1,\xi _2)$ as the central surface of a shell we define a finite thickness shell as the following space domain
\begin{equation} \label{eq:3D-surface}
\vec{r}(\xi _1,\xi _2,\xi _3) = \vec{\varsigma }(\xi _1,\xi _2) + \xi _3 \vec{e}_3,
\end{equation}
where $\xi _3\in[-h/2,h/2]$ is the cross section coordinates with $h$ being the shell thickness. Similarly to the previous section we use the assumption that the thickness $h$ is infinitesimally small and suppose that the magnetisation does not depend on $\xi _3$, hence $\vec{m} = \vec{m}(\xi _1,\xi _2)$.

In order to calculate the exchange energy of the shell, we start with the definition \eqref{eq:Eex-Cartesian} and substitute the Cartesian components of the magnetisation vector $m_i$ in terms of the curvilinear components $m_\alpha$ as follows $m_i=m_\alpha  \left({\vec e}_\alpha \cdot\hat{\vec x}_i\right)$. By applying del operator in its curvilinear form $\vec\nabla\equiv\vec{e}_\alpha \nabla_\alpha  \equiv{\vec e}_\alpha (g_{\alpha \alpha })^{-1/2}\partial_\alpha$, we get the exchange energy in the form, similar to \eqref{eq:E-1D}:
\begin{subequations} \label{eq:E-2D}
\begin{equation} \label{eq:E-sum-2D}
\mathscr{E}_{\mathrm{ex}} = \nabla_\alpha  \vec{m}\cdot \nabla_\alpha  \vec{m} = \mathscr{E}_{\mathrm{ex}}^{0}  + \mathscr{E}_{\mathrm{ex}}^{A} + \mathscr{E}_{\mathrm{ex}}^{D}.
\end{equation}
The first term is the isotropic part of exchange expression
\begin{equation} \label{eq:E-2D-0}
\mathscr{E}_{\mathrm{ex}}^{0} = \vec{\nabla}m_{\alpha }\cdot \vec{\nabla}m_{\alpha },
\end{equation}
which has formally the same form as for the plane surface.

Similarly to 1D case, the curvature manifests itself in the effective anisotropy--like term $\mathscr{E}_{\mathrm{ex}}^{A}$ and the effective Dzyaloshinskii interaction term $\mathscr{E}_{\mathrm{ex}}^{D}$ as follows:
\begin{equation} \label{eq:E-2D-A-D}
\eqalign{
\mathscr{E}_{\mathrm{ex}}^{A} = K_{\alpha \beta } m_\alpha  m_\beta ,  \qquad &K_{\alpha \beta } = \nabla_\gamma  \vec{e}_\alpha  \cdot \nabla_\gamma  \vec{e}_\beta ,\\
\label{eq:E-2D-D} %
\mathscr{E}_{\mathrm{ex}}^{D} = 2D_{\alpha \beta \gamma } m_\beta  \nabla_\gamma  m_\alpha ,  \qquad &D_{\alpha \beta \gamma } = \vec{e}_\alpha  \cdot \nabla_\gamma  \vec{e}_\beta .
}
\end{equation}
\end{subequations}
By using Gauss--Codazzi equation \eqref{eq:Gauss}, one can show that the components of the tensor $K_{\alpha \beta }$ have a bilinear form with respect to the components of the second fundamental form $b_{\mu \nu }$, it emulates a `geometrical potential' closely related to the potential which arises in the quantum mechanical problem of the particle rigidly bounded to a surface \cite{Costa81}. The effective Dzyaloshinskii interaction coefficients $D_{\alpha \beta \gamma }$ are linear with respect to the $b_{\mu \nu }$ components. This effective interaction is a source of possible magnetochiral effects, such as the vortex polarity--chirality coupling \cite{Kravchuk12a}, and interrelation between chiralities of the sample and its magnetisation subsystem for M\"obius rings \cite{Pylypovskyi14c}. The effects of curvature induced magnetochirality were reviewed recently in Ref.~\cite{Hertel13a}.

Let us use the angular parametrisation \eqref{eq:theta-phi-1D} with $\theta=\theta (\xi_1,\xi_2)$ being the colatitude and $\phi=\phi (\xi_1,\xi_2)$ being the azimuthal angle in the local frame of reference. In terms of $\theta $ and $\phi $ the exchange energy density $\mathscr{E}_{ex}$ reads
\begin{equation} \label{eq:E-2D-final}
\!\!\mathscr{E}_{\mathrm{ex}}^{2d}\!\! = \left[\vec{\nabla}\theta  \!\! - \vec{\varGamma }(\phi )\right]^2\!\! + \left[\sin\theta \left(\vec{\nabla}\!\phi-\vec{\varOmega }\right)\! - \!\cos \theta \frac{\partial \vec{\varGamma }(\phi )}{\partial\phi }\right]^2\!\!\!\!\!.
\end{equation}
Here the vector $\vec\varOmega$ is a modified spin connection, $\vec\varOmega=\vec e_\mu (\vec{e}_1\cdot \nabla_\mu  \vec{e}_2)$, and vector $\vec\varGamma$ is determined as follows
\begin{equation} \label{eq:Gamma-def}
\vec{\varGamma }(\phi ) \!=\!||H_{\alpha \beta }||\vec{\varepsilon }(\phi )=\mathcal{H}\, \vec{\varepsilon }(\phi ) + \sqrt{\mathcal{H}^2-\mathcal{K}}\, \vec{\varepsilon }(\upsilon -\phi ),
\end{equation}
where $\vec{\varepsilon }(\phi )=\left(\cos\phi,\sin\phi \right)$  and $\tan\upsilon=2H_{12}/(H_{11}-H_{22})$. Very recently we derived the exchange energy for the curved shell in the form \eqref{eq:E-2D-final} in Ref.~\cite{Gaididei14}.

It is instructive to establish a link between the 2D energy \eqref{eq:E-2D-final} and the 1D expression \eqref{eq:E-1D-final}. For this purpose we define the surface $\vec\varsigma (\xi_1,\xi_2)$ as a local extension of the curve $\vec\gamma (s)$ in the following way
\begin{equation} \label{eq:surf-gamma}
\vec{\varsigma }(\xi _1\equiv s,\xi _2) = \vec{\gamma }(s)+ \xi _2 \vec{e}_2(s).
\end{equation}
By using the Frenet–-Serret formulae \eqref{eq:1D-basis} and \eqref{eq:Frenet–Serret} one can easily find the corresponding metric tensor and the Hessian matrix. For points of the curve $\vec\gamma$ one has
\begin{equation} \label{eq:g-n-h}
\lim\limits_{\xi_2\to0} \left\|g_{\mu \nu }\right\| = \mathrm{diag}(1,1), \quad \lim\limits_{\xi_2\to0} \left\|H_{\mu \nu }\right\| = \mathrm{adiag}(\tau ,\tau ).
\end{equation}
According to \eqref{eq:Gamma-def} one can find that $\vec\varGamma(\vec\gamma )  = \vec e_1\tau \sin \phi $ and the spin connection $\vec{\varOmega }(\vec\gamma ) = -\kappa \vec e_1$. Assuming now that the magnetisation on the surface \eqref{eq:surf-gamma} depends on $s$ only, one obtains $\vec\nabla\theta=\theta'(s)\vec e_1$ and $\vec\nabla\phi=\phi'(s)\vec e_1$, and finally the 2D energy \eqref{eq:E-2D-final} takes the form \eqref{eq:E-1D-final}.

Let us take into account the anisotropy term, starting from the energy functional \eqref{eq:E-total} and choosing the anisotropy axis $\vec{n}=\vec{e}_3$, i.e. along the normal to the surface. Then the total energy density of the shell is
\begin{equation} \label{eq:Eex+Ea-2D}
\mathscr{E}^{2d} =  \ell^2\mathscr{E}_{\mathrm{ex}}^{2d}+ \lambda \cos^2\theta .
\end{equation}
Similarly to the case of 1D nanowire the general static solutions for the highly anisotropic 2D shell can be obtained. As previously, this solution can be treated as a result of acting of an effective curvature induced magnetic field. Since this approach for the 2D case was already discussed in Ref.~\cite{Gaididei14}, we limit ourselves with an example for positive anisotropy constant, namely $\lambda \gg1$. This corresponds to strong easy-surface anisotropy. In this case the magnetisation has a  quasitangential distribution: $\theta=\pi/2+\vartheta $ with $|\vartheta |\ll1$ and the total energy density reads \cite{Gaididei14}
\begin{equation} \label{eq:En-easy-surf-all}
%\fl
\eqalign{
&\mathscr{E}^{2d}\approx \mathscr{E}^\mathrm{t}+F\vartheta +\lambda \vartheta ^2,\\
&\mathscr{E}^{\mathrm{t}}=\ell^2 \left[\vec{\varGamma }^2+(\vec\nabla\phi-\vec\varOmega )^2 \right],\quad
F=2\ell^2\left[\nabla\cdot\vec\varGamma + (\nabla\phi-\vec\varOmega )\frac{\partial\vec\varGamma }{\partial\phi } \right]\!\!,
}
\end{equation}
where $\mathscr{E}^{\mathrm{t}}$ is the energy density of the strictly tangential distribution and $F(\phi )$ can be treated as the amplitude of an effective curvature induced magnetic field oriented along vector $\vec\varepsilon $. Minimisation of \eqref{eq:En-easy-surf-all} with respect to $\vartheta $ and $\phi$ results in
\begin{equation} \label{eq:theta-surf}
\theta  = \frac{\pi }{2} - \frac{1}{2\lambda } F(\phi )+\mathcal{O}\left(\frac{1}{|\lambda |^2}\right),
\end{equation}
where the equilibrium function $\phi$ is obtained as a solution of the equation $\delta \mathscr{E}^\mathrm{t}/\delta \phi=0$. Accordingly to \eqref{eq:theta-surf} the strictly tangential solution is realised only for a specific case $F(\phi )\equiv0$. Consideration of the case of strong easy-normal anisotropy ($\lambda \ll-1$) can be found in Ref.~\cite{Gaididei14}.

{It is worth noticing here that the appearance of effective Dzyaloshinskii-like interaction (or in other words, Lifshitz invariants) in curved magnetic systems is inherently coupled with the fact that in contrast to the order parameter $\vec{u}$ in  the Ginzburg--Landau functional given by Eq. \eqref{eq:GL}, the magnetisation $\vec{m}$ is a vector which is transformed by transformations of the real space. Formally it is expressed in Eq. \eqref{eq:m-Cartesian} which shows that the coupling  between the Cartesian and curvilinear components of the magnetisation vector is space dependent and therefore the action of the del operator on the corresponding coefficients cannot be ignored.

\section{1D example: ground state and magnon spectrum for ring nanowire}
\label{sec:ring}

%==================================================================\
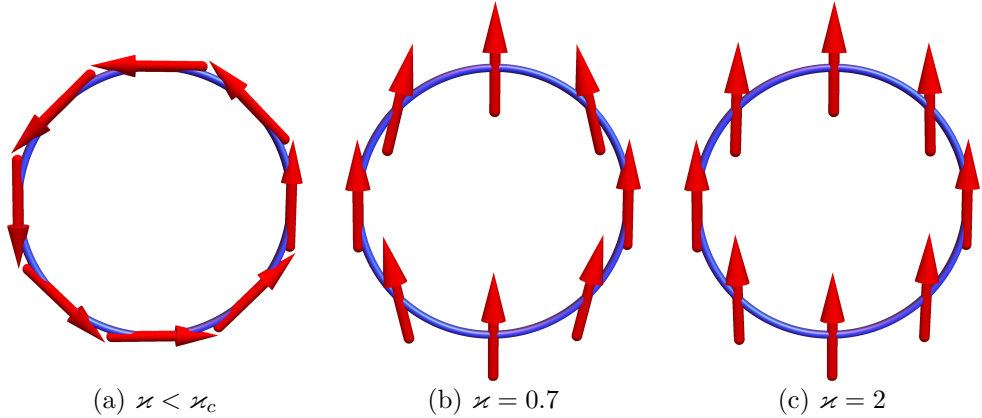
\begin{figure}
\begin{center}
\begin{tikzpicture}%[scale=1,show grid={left,above,true}]
\node[above] at (0,0.7) {\includegraphics[width=0.3\textwidth]{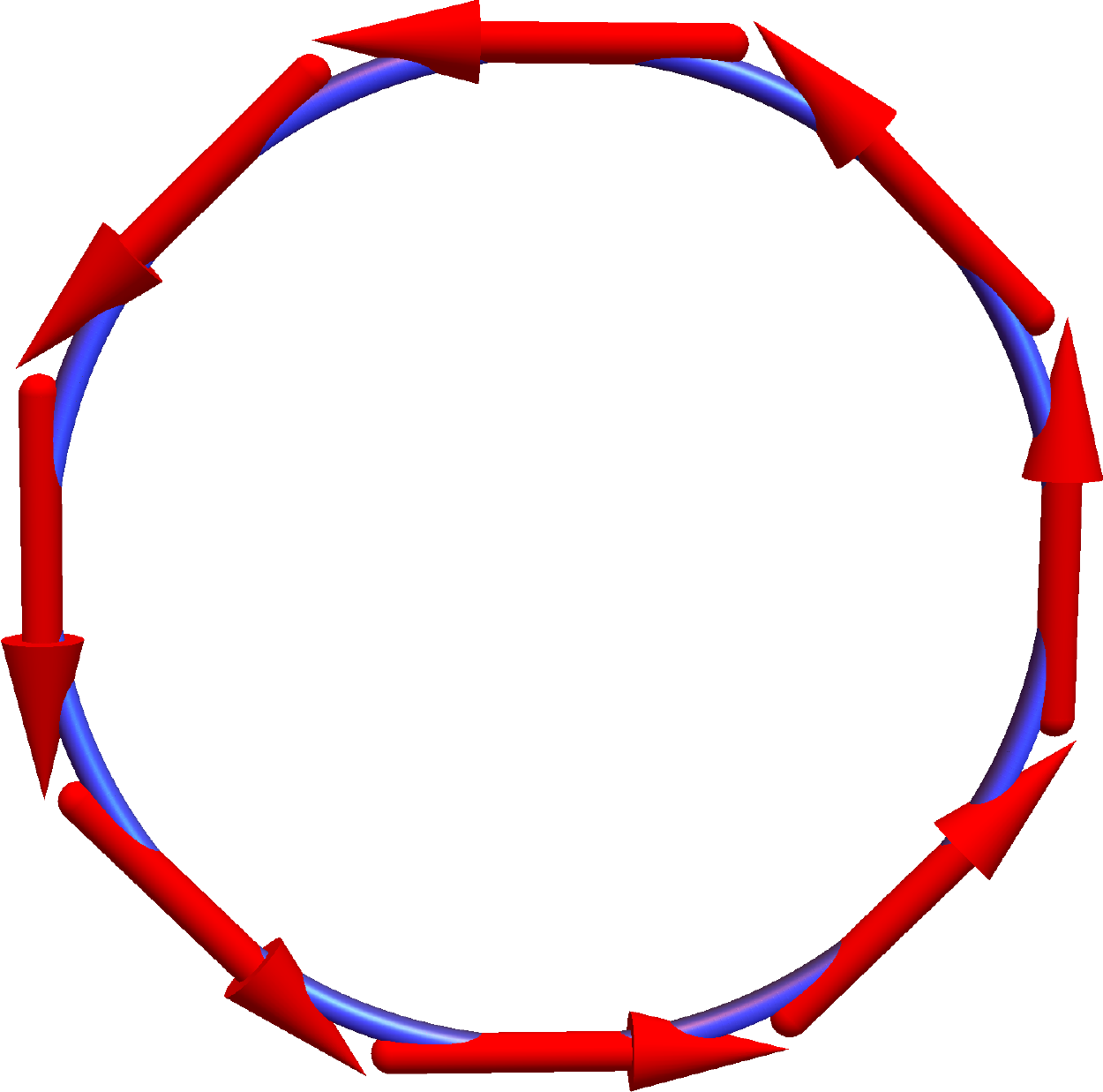}};
\node[above] at (4.5,0.25) {\includegraphics[width=0.3\textwidth]{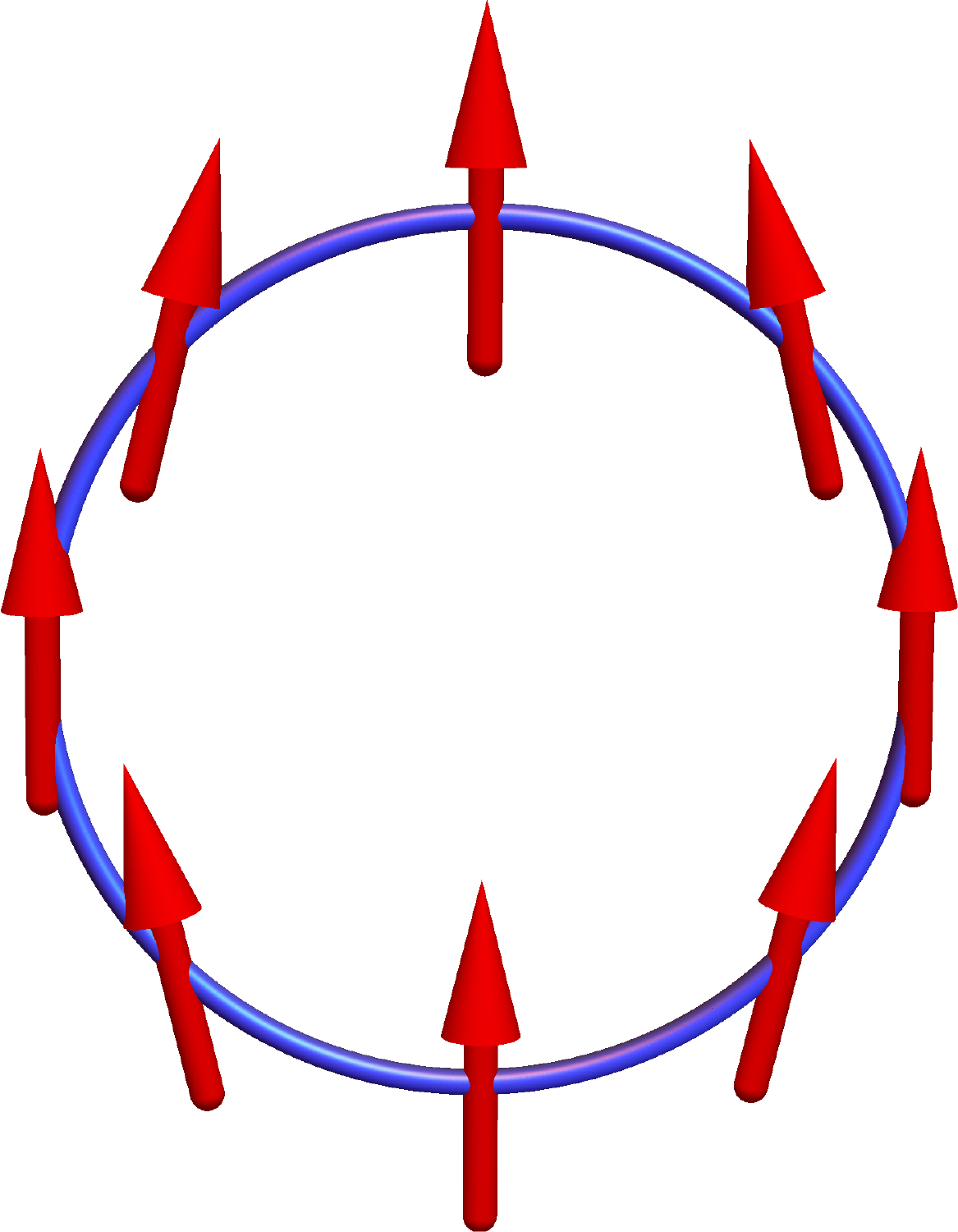}};
\node[above] at (9,0.25) {\includegraphics[width=0.3\textwidth]{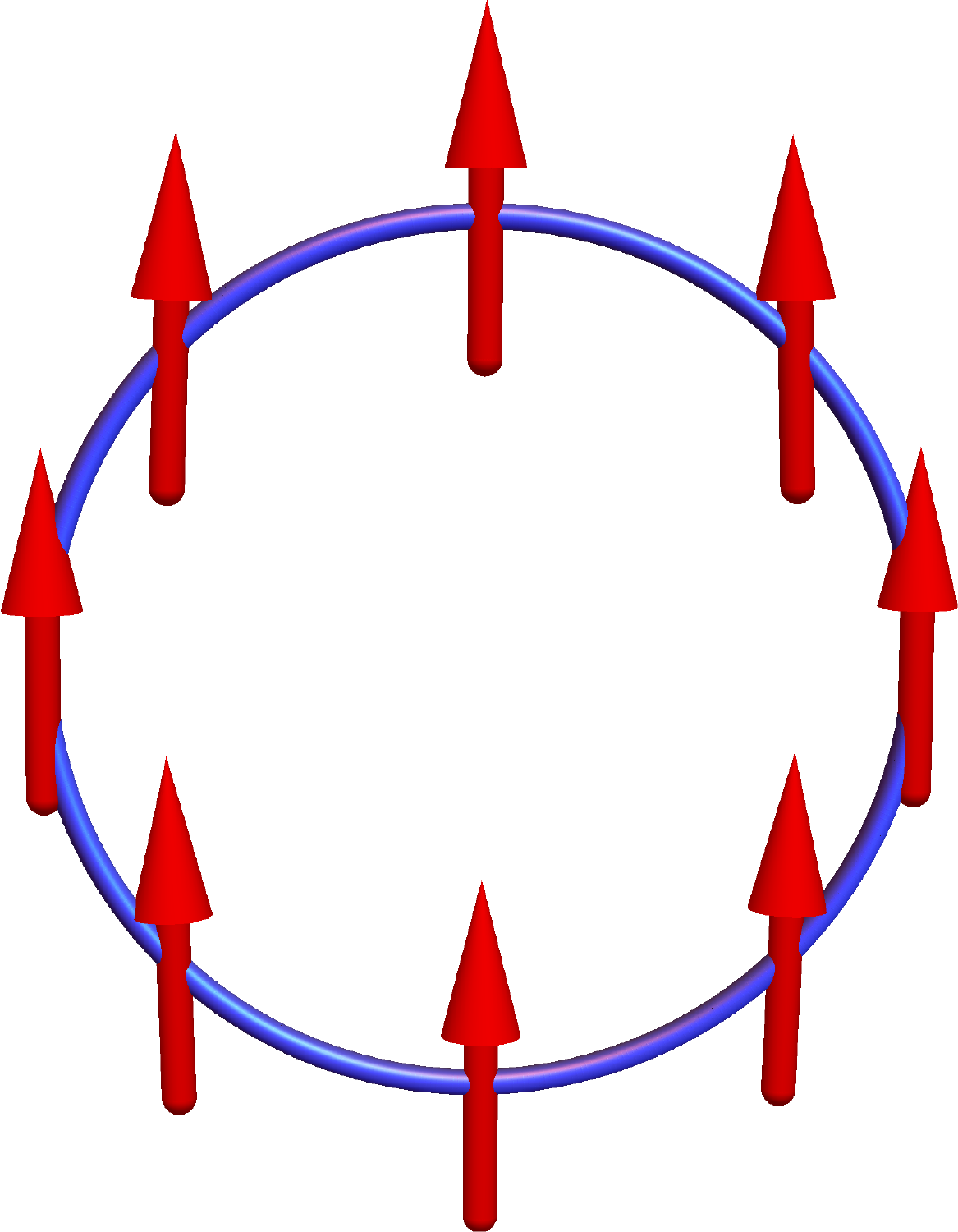}};
\node[above] at (0.0,-0.2) {(a) $\varkappa <\varkappa _c$};
\node[above] at (4.5,-0.2) {(b) $\varkappa =0.7$};
\node[above] at (9.0,-0.2) {(c) $\varkappa =2$};
\end{tikzpicture}
%(a) $\varkappa <\varkappa _c$ \hfill (b) $\varkappa =0.7$ \hfill (c) $\varkappa =2$
\end{center}
\caption{Magnetisation distribution of the ground state in a ring wire with different reduced curvatures $\varkappa $: (a) vortex state, (b) and (c) onion states \eqref{eq:onion}.}
\label{fig:ring}
\end{figure}
%==================================================================/

As an example of our approach for curved magnets we consider the magnetisation distribution in a simplest curvilinear system with the constant curvature and no torsion, i.e. a ring-shaped wire (circumference). Using the arc length coordinate $s$, we put $\vec{\gamma }(s) = \left\{\kappa ^{-1} \cos(\kappa  s),\kappa ^{-1} \sin(\kappa  s),0\right\}$. Let us consider the case with easy tangential anisotropy $\lambda <0$. The total energy \eqref{eq:E-total} of the wire with the area cross-section $\mathcal{S}$ has the form $E=2\pi \mathcal{S}|\lambda |\kappa ^{-1}\mathcal{E}$ where
\begin{equation} \label{eq:E-ring}
\mathcal{E}= \frac{1}{2\pi } \int\limits_0^{2\pi }\left\{\varkappa ^2\left[(\partial_\chi  \theta )^2 + \sin^2\theta (1-\partial_\chi \Phi )^2\right] - \sin^2\theta  \sin^2\Phi \right\}\mathrm{d}\chi .
\end{equation}
Here we used the modified magnetisation azimuthal angle $\Phi =\pi /2-\phi $, the angular variable $\chi  \equiv \kappa  s$, the reduced curvature $\varkappa  \equiv \kappa  w$ and the `magnetic' length $w = \ell/\sqrt{|\lambda |}$. The minimisation of the energy \eqref{eq:E-ring} results in $\theta =\pi /2$ and the azimuthal angle $\Phi $, which satisfies the pendulum equation
\begin{equation} \label{eq:pendulum}
\varkappa ^2 \partial_{\chi \chi } \Phi  + \sin\Phi \cos\Phi =0.
\end{equation}
The homogeneous (in the curvilinear reference frame) solution corresponds to the planar vortex state:
\begin{equation} \label{eq:vortex}
\Phi ^{\mathrm{vor}}=\mathcal{C}\frac{\pi }{2},  \qquad \theta ^{\mathrm{vor}}=\frac{\pi }{2},
\end{equation}
which is well known for the magnetic nanorings \cite{Klaui03a,Guimaraes09}, therefore we name it a \emph{vortex} solution, the parameter $\mathcal{C}=\pm1$ is called the vortex chirality (clockwise or counterclockwise). The energy of the vortex state $\mathcal{E}^{\mathrm{vor}} = - 1+\varkappa ^2$.

Another, inhomogeneous solution of the pendulum equation \eqref{eq:pendulum} reads
\begin{subequations} \label{eq:onion}
\begin{equation} \label{eq:phi-sol}
\Phi ^{\mathrm{on}}(\chi ) = \mathrm{am}(x,k),\qquad x=\frac{2\chi }{\pi }\mathrm{K}(k), \qquad \theta ^{\mathrm{on}}=\frac{\pi }{2},
\end{equation}
where $\mathrm{am}(x,k)$ is Jacobi's amplitude \cite{NIST10} and the modulus $k$ is determined by the condition
\begin{equation} \label{eq:k-sol}
2\varkappa  k\mathrm{K}(k)=\pi ,
\end{equation}
\end{subequations}
with $\mathrm{K}(k)$ being the complete elliptic integral of the first kind \cite{NIST10}. The corresponding magnetisation solution is analogous to a well-known onion sate \cite{Klaui03a,Guimaraes09} typical for the ring geometry, hence we refer \eqref{eq:onion} as to the \emph{onion} state. The energy of the onion state reads
\begin{equation} \label{eq:E-onion}
\mathcal{E}^{\mathrm{on}} =  \frac{4 \varkappa }{\pi  k}\mathrm{E}(k) - \varkappa ^2 - \frac{1}{k^2},
\end{equation}
where $\mathrm{E}(k)$ is the complete elliptic integral of the second kind \cite{NIST10}. The equality of energies $\mathcal{E}^{\mathrm{vor}} = \mathcal{E}^{\mathrm{on}}$ determines the critical curvature $\varkappa _c\approx 0.657$, which separates the vortex state ($\varkappa <\varkappa _c$) and the onion one ($\varkappa >\varkappa _c$). The typical magnetisation distribution is shown in  Fig.~\ref{fig:ring}.

To analyse the magnons in the system we linearise the Landau--Lifshitz equations \eqref{eq:LL-angular} on the background of $\theta _0=\pi /2$ and $\Phi _0(\chi )$, which corresponds to the vortex state ($\Phi _0=\Phi ^{\mathrm{vor}}$) or the onion one ($\Phi _0=\Phi ^{\mathrm{on}}$) depending on the curvature $\varkappa $. For the small deviations $\vartheta =\theta -\theta _0$ and $\varphi =\Phi -\Phi _0(\chi )$ we get the set of linear equations:
\begin{subequations} \label{eq:LL-linearised-eqs}
\begin{equation} \label{eq:LL-linearised}
%\fl
\left[-\varkappa ^2\partial_{\chi \chi } + V_1(\chi )\right]\vartheta  = -\partial_\tau  \varphi ,\quad \left[-\varkappa ^2\partial_{\chi \chi } + V_2(\chi )\right]\varphi  = \partial_\tau  \vartheta ,
\end{equation}
where $\partial_\tau $ is the derivative with respect to the dimensionless time $\tau =\Omega _0 t$ with $\Omega _0=2\omega _0|\lambda |$. Here the ``potentials'' $V_1(\chi )$ and $V_2(\chi )$ are as follows:
\begin{equation} \label{eq:potentials}
%\fl
V_1(\chi ) = \sin^2\Phi _0 - \varkappa ^2\left[1-\partial_\chi  \Phi _0(\chi )\right]^2,\qquad V_2(\chi ) = - \cos 2\Phi _0(\chi ).
\end{equation}
\end{subequations}
We apply the partial wave expansion
\begin{equation} \label{eq:partial-wave}
\fl
\vartheta (\chi ,\tau ) = \sum_{m=0}^\infty\!\! \vartheta _m\cos(m\chi -\Omega \tau +\delta _m), \quad \varphi (\chi ,\tau ) = \sum_{m=0}^\infty\!\! \varphi _m\sin(m\chi -\Omega \tau +\delta _m)
\end{equation}
with $m$ being the azimuthal quantum numbers, $\delta _m$ being arbitrary phases, and $\Omega  = \omega /\Omega _0$ being dimensionless frequencies. Let us mention that Eqs.~\eqref{eq:LL-linearised-eqs} for the partial waves $\vartheta _m$ and $\varphi _m$ are invariant under the conjugation $\Omega \to-\Omega $, $m\to-m$, $\delta _m\to-\delta _m$, $\vartheta _m\to\vartheta _m$, and $\varphi _m\to-\varphi _m$. In classical theory we can choose any sign of frequency; nevertheless, to make a contact with a quantum mechanics with a positive frequency and energy $\mathscr{E}_k=\hslash \omega _k$, we discuss the case $\Omega >0$ only.

First we consider the magnons on the background of the vortex state \eqref{eq:vortex}. In this case $V_1=1-\varkappa ^2$ and $V_2=1$. By substituting the expansion \eqref{eq:partial-wave} into Eqs.~\eqref{eq:LL-linearised-eqs} one can calculate the following spectrum of magnon eigenstates:
\begin{equation} \label{eq:spectrum-vortex}
\Omega _m^{\mathrm{vor}}(\varkappa ) = \sqrt{\left( 1+ \varkappa ^2 m^2\right)\left( 1+ \varkappa ^2 m^2-\varkappa ^2\right)}.
\end{equation}
The lower eigenfrequencies are plotted in the Fig.~\ref{fig:circle_spectrum}.

In the limit case of a quasi-straight wire ($\varkappa \to0$) the magnon frequencies read
\begin{equation*}
\Omega _m^{\mathrm{vor}}(\varkappa ) = 1-\frac{\varkappa ^2}{2}+\varkappa ^2m^2+\Or \left(\varkappa ^4\right).
\end{equation*}
Thus the curvature decreases the gap as compared to the case of the straight wire $(\varkappa =0)$  with dispersion $\Omega _s(\mathfrak{K}) = 1+ \mathfrak{K}^2$, where $\mathfrak{K}=\varkappa m$ is the corresponding normalised wave vector.

%==================================================================\
\begin{figure}
\begin{center}
\includegraphics[width=0.75\columnwidth]{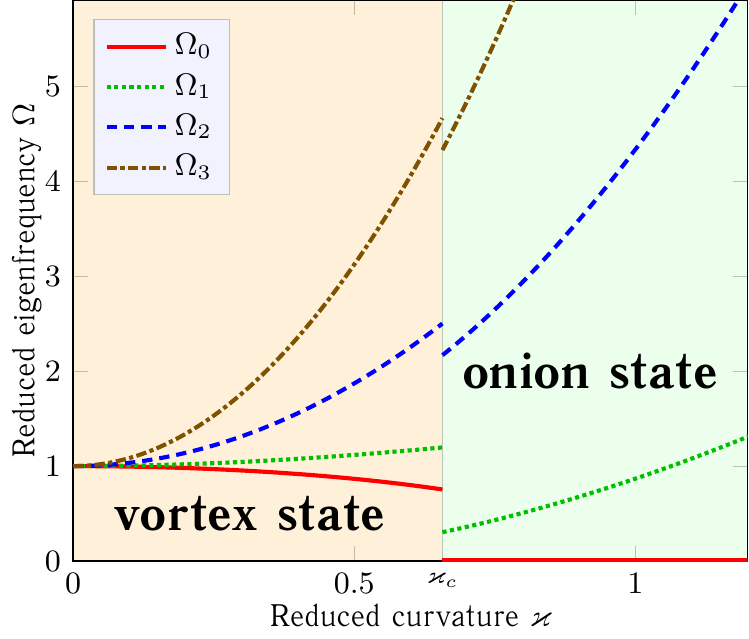}
\end{center}
\caption{The lowest eigenfrequencies of linear excitations in a ring nanowire depending on the curvature $\varkappa $.}
\label{fig:circle_spectrum}
\end{figure}
%==================================================================/

Let us consider now the magnons on background of the onion state \eqref{eq:onion}. By substituting $\Phi _0=\Phi ^\mathrm{on}$ into \eqref{eq:potentials} one can present the potentials $V_1(\chi )$ and $V_2(\chi )$ as the following Fourier expansions \cite{NIST10}
\begin{equation} \label{eq:V12-series}
\eqalign{
V_1=A_0+\sum\limits_{n=1}^\infty A_n\cos(2n\chi ),\quad V_2=B_0+\sum\limits_{n=1}^\infty B_n\cos(2n\chi ),\\
A_0=\frac{1}{k^2}-\frac{4}{\pi }\frac{\varkappa }{k}\mathrm{E}(k)+\varkappa ^2,\quad A_n=8\varkappa ^2q^n\left[\frac{1}{1+q^{2n}}-\frac{2n}{1-q^{2n}}\right],\\
B_0=\frac{2}{k^2}-\frac{4}{\pi }\frac{\varkappa }{k}\mathrm{E}(k)-1,\quad B_n=-16\varkappa ^2\frac{nq^n}{1-q^{2n}}.
}
\end{equation}
where Jacobi's nome $q$ is given in terms of the modulus $k$ by $q=\exp\left(-\pi \mathrm{K}(\sqrt{1-k^2})/\mathrm{K}(k)\right)$ \cite{NIST10}. At critical point $\varkappa _c$ the nome $q(\varkappa _c)\approx 0.135$, and its value rapidly tends to zero with $\varkappa $. Thus we can restrict ourselves with few lower Fourier harmonics.

Now by substituting \eqref{eq:partial-wave} and \eqref{eq:V12-series} into \eqref{eq:LL-linearised} and multiplying the Fourier series, we get the following set of equations
\begin{equation} \label{eq:phi-theta-onion}
\eqalign{
&(\varkappa ^2m^2+A_0)\vartheta _m+\frac12\sum\limits_{n=1}^\infty A_n(\vartheta _{m+2n}+\vartheta _{m-2n}) = \Omega  \varphi _m,\\
&(\varkappa ^2m^2+B_0)\varphi _m+\frac12\sum\limits_{n=1}^\infty B_n(\varphi _{m+2n}+\varphi _{m-2n}) = \Omega  \vartheta _m,
}
\end{equation}
where the conventional rule $f_{-|n|}=f_{|n|}$ is used for the amplitudes $\vartheta _{n}$ and $\varphi _{n}$.

We do not possess the exact solution of the infinite set of equations \eqref{eq:phi-theta-onion}. As a first approach, by neglecting the modes coupling one obtains
\begin{equation} \label{eq:Omega-no-coupl}
\Omega _m^{(u)} = \sqrt{(\varkappa ^2m^2+A_0)(\varkappa ^2m^2+B_0)}.
\end{equation}
The coupling results in the mixing of different partial waves. However the influence of coupling decreases with $n$ due to the rapid decay of $A_n$ and $B_n$, hence \eqref{eq:Omega-no-coupl} provides good enough estimation of frequencies for not very small azimuthal quantum number $m$.

An exception is $\Omega =0$: in this case the zero (Goldstone) mode is realised due to arbitrary direction of the onion axis. This eigenstate has the following form
\begin{equation} \label{eq:phi-Goldstone}
\varphi ^{\mathrm{G}}(\chi ) = \partial_\chi  \Phi ^{\mathrm{on}}(\chi ) \propto \mathrm{dn}(x,k),\qquad \vartheta ^{\mathrm{G}}(\chi )=0,  \qquad \Omega ^{\mathrm{G}}=0.
\end{equation}
Using the Fourier expansion of Jacobi's function $\mathrm{dn}(x,k)$, one can easily see that the Goldstone mode $\varphi ^{\mathrm{G}}(\chi )$ contains infinite number of partial waves, hence the coupling between different partial waves for this mode is crucial. One has to stress that as distinct from the vortex case, eigenstates on the background of the onion state do not coincide with partial waves: each eigenstate with eigenfrequency $\Omega _n$ corresponds to a set of partial waves with different azimuthal quantum numbers $m$ due to the coupling. The lowest eigenfrequencies, calculated using \eqref{eq:phi-theta-onion} with account of only four lowest partial waves $\Omega _n$, are plotted in the Fig.~\ref{fig:circle_spectrum}.

The spectrum of the narrow nanorings is well studied experimentally \cite{Giesen07,Demokritov09}. It should be noted that the typical for experiments ring radii $R$ are about hundreds of nanometres, while the typical magnetic length $w$ is about ten nanometres, hence the dimensionless curvature $\varkappa \approx w/R\ll1$. That is why in most of experiments the ground state of the ring is the vortex state and the onion one appears only under influence of external magnetic field \cite{Klaui03a}.

One has to stress that we do not discuss here the influence of the dipolar interaction on the magnetisation structure supposing that the thickness is much smaller than the exchange length. Nevertheless it is instructive to compare our results for the critical curvature $\varkappa _c$ with the boundary between different phases in magnetic rings. Our case of circumference--wire corresponds to the very narrow ring. It is well known \cite{Kravchuk07} that depending on the geometrical and magnetic parameters of the nanoring, there exist different magnetic phases in magnetically soft ring: easy-axis, easy-plane, and planar vortex phases. The lowest bound for the vortex state magnetic ring is given by the triple point $R^{\mathrm{(tr)}}\approx \ell \sqrt{3}$ for the infinitesimally narrow ring  \cite{Kravchuk07}. For rough estimation of the critical curvature we can simply replace the magnetic length $w$ by the exchange length $\ell$, hence $\varkappa \approx \ell/R^{\mathrm{(tr)}} = 1/\sqrt{3}\approx0.577$, which is close to $\varkappa _c\approx 0.657$. One has to note that the monodomain state in \cite{Kravchuk07} was supposed to be the easy-axial one instead of the onion state.

\section{2D example: linear magnetisation dynamics for a cone shell}
\label{sec:cone}

In this section we illustrate our approach for curved shells considering a cone shell with high easy--surface anisotropy. Recently we found out the ground magnetisation states of side surface of a right circular truncated cone \cite{Gaididei14}. In the current study we solve the dynamical problem of spin waves on the background of the ground state.

Let us consider the cone surface, where the radius of the truncation face is $R$ and the length of the cone generatrix is $L$. Varying the generatrix inclination angle $\psi \in [0,\pi/2]$ one can continuously proceed from the planar ring ($\psi =0$) to the cylinder surface ($\psi =\pi /2$), for notations see inset (a) in Fig.~\ref{fig:cone-spectrum}. The cone surface can be parameterised as follows,
\begin{equation} \label{eq:cone-param}
x+i y =(R+\xi _2\cos\psi )\exp(\imath \xi _1),\qquad z=\xi _2\sin \psi ,
\end{equation}
with the curvilinear coordinates $\xi _1\in S^1$ and $\xi _2\in[0,L]$. The parametrisation \eqref{eq:cone-param} generates the following geometrical properties: the metric tensor $\left\|g_{\alpha \beta } \right\| = \mathrm{diag}(g;1)$, the modified spin connection ${\vec \varOmega}={\vec e}_1\cos\psi /\sqrt{g}$, and $\vec \varGamma =-{\vec e}_1 \sin \psi \cos \phi /\sqrt{g}$, where $\sqrt g=R+ \xi _2\cos\psi $ \cite{Gaididei14}. As above, we use here the angular parametrisation \eqref{eq:theta-phi-1D} for the magnetisation.

We limit ourselves by the case of the strong easy-surface anisotropy. In accordance to \eqref{eq:theta-surf}, the magnetisation polar angle $\theta \approx \pi /2$. Similar to 1D case, the azimuthal magnetisation angle $\phi $ satisfies the pendulum equation \cite{Gaididei14}, cf. \eqref{eq:pendulum}
\begin{equation} \label{eq:pendulum-cone}
\partial_{\xi _1\xi _1} \phi  +\sin^2\psi  \sin\phi \cos\phi =0.
\end{equation}
The ground state of such a cone is the onion state $\phi ^{\mathrm{on}}$  for $\psi <\psi _c\approx 0.8741$ and axial one $\phi ^{\mathrm{ax}}$ for $\psi >\psi _c$ \cite{Gaididei14}:
\begin{equation} \label{eq:cone-gs}
\phi ^{\mathrm{on}}(\xi _1) = \mathrm{am}(x,k),\quad x=\frac{2\xi _1}{\pi }\mathrm{K}(k), \qquad \phi ^{\mathrm{ax}}=\pm \pi /2,
\end{equation}
where the modulus $k$ of the Jacobi's amplitude is determined by the condition $2k\mathrm{K}(k)=\pi \sin\psi$, cf. \eqref{eq:onion}. The magnetisation dynamics follows the Landau--Lifshitz equation \eqref{eq:LL-angular}. In the case of a high easy-surface anisotropy $\lambda \gg\ell^2/\mathcal{R}^2$, one can derive the dynamical equation for the in--surface magnetisation angle $\phi $:
\begin{equation} \label{eq:phi-dynamics}
\frac{\partial_{tt}\phi }{4\lambda \omega _0^2\ell^2} = \nabla\cdot(\nabla\phi-\vec\varOmega) - \vec\varGamma \cdot\frac{\partial\vec\varGamma }{\partial\phi }.
\end{equation}
For the cone surface \eqref{eq:cone-param} the dynamic equation \eqref{eq:phi-dynamics} takes the form
\begin{equation} \label{eq:dynamic}
\frac{g}{4\lambda \omega _0^2\ell^2}\partial_{tt} \phi = \partial_{\xi _1\xi _1}\phi + g \partial_{\xi _2\xi _2} \phi + \sqrt{g}\cos\psi \partial_{\xi _2}\phi + \frac12 \sin^2\psi \sin2\phi.
\end{equation}
Now after linearising this equation on the background of the onion state $\phi ^{\mathrm{on}}(\xi _1)$, we can present the small deviation $\varphi =\phi -\phi ^{\mathrm{on}}$ in the following form:
\begin{subequations} \label{eq:Lame-Bessel}
\begin{equation} \label{eq:linear-sep}
\varphi (\xi _1,\xi _2,t) = e^{i\omega t}\mathrm{P}(\rho )\mathrm{X}(x),
\end{equation}
where $\rho=1+\frac{\xi _2}{R}\cos\psi$ and $x$ is defined in \eqref{eq:cone-gs}. By separating variables one can find that the angular part $\mathrm{X}(x)$ satisfies the Lam\'{e} equation \cite{NIST10}
\begin{equation} \label{eq:Lame}
\mathrm{X}'' + \left[\Lambda -2k^2\mathrm{sn}^2(x,k) \right]\mathrm{X}=0,
\end{equation}
where $\mathrm{sn}(x,k)$ is a Jacobi elliptic function \cite{NIST10}. The periodic solution of \eqref{eq:Lame} which corresponds to the lowest eigenvalue $\Lambda =k^2$ \cite{NIST10} coincides (up to the constant) with the following Lam\'{e} function $\mathrm{X}(x) = \mathcal{C}\,\mathrm{Ec}_1^{0}(x,k^2)$. Then the function $\mathrm{P}(\rho )$ appears as the solution $\mathrm{P}(\rho )= \mathcal{C}_1 \mathrm{J}_0(\mathfrak{q}\rho ) + \mathcal{C}_2 \mathrm{N}_0(\mathfrak{q}\rho )$ of a zero-order Bessel equation, where $\mathfrak{q}=\Omega /\cos\psi $ with $\Omega  = \omega /\omega _c$ and $\omega _c=2\omega _0\sqrt{\lambda }\ell/R$. Using the boundary conditions
\begin{equation}\label{eq:bc}
\mathrm{P}'(0)= \mathrm{P}'(\rho_0) = 0,
\end{equation}
where $\rho_0=1+\frac{L}{R}\cos\psi$ one can determine the eigenvalues from the following equation $\mathrm{J}_1(\mathfrak{q}) \mathrm{N}_1(\mathfrak{q}\rho_0) = \mathrm{J}_1(\mathfrak{q}\rho _0) \mathrm{N}_1(\mathfrak{q})$, whose numerical solution is plotted in  Fig.~\ref{fig:cone-spectrum} for the case $\psi <\psi _c$.
\end{subequations}

Similar to \eqref{eq:phi-Goldstone}, there is the zero (Goldstone) mode for the magnon oscillations on the onion background. The eigenstate for the zero mode reads:
\begin{equation} \label{eq:X-Goldstone}
\varphi ^{\mathrm{G}}(\xi _1) = X^{\mathrm{G}}(x) = \mathrm{dn}(x,k),  \qquad \Lambda ^{\mathrm{G}}=k^2, \qquad \Omega ^{\mathrm{G}}=0.
\end{equation}

%==================================================================\
\begin{figure}
\begin{center}
\includegraphics[width=0.9\textwidth]{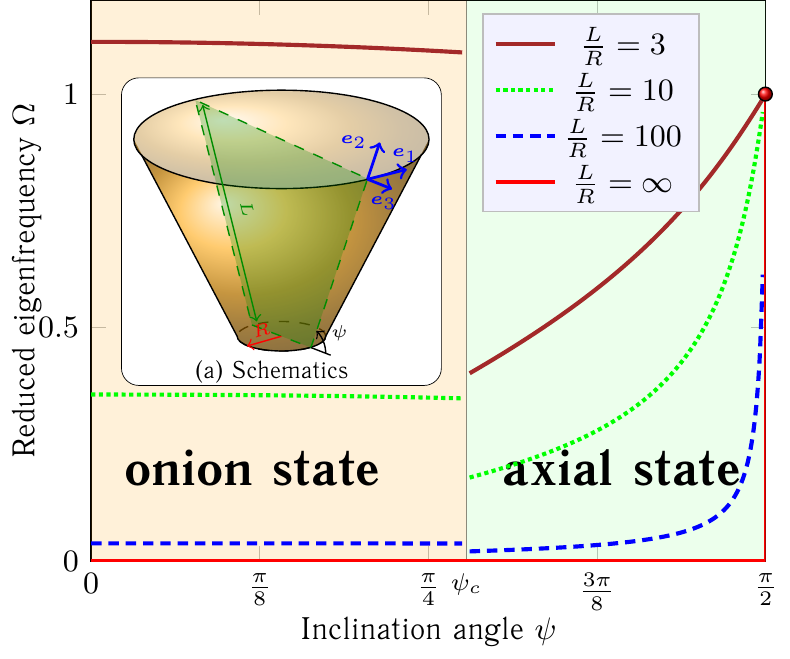}
\end{center}
\caption{The lowest frequencies of linear excitations over the easy-surface ground states of the cone depending on the relative generatrix length $L/R$ and inclination angle $\psi $. Inset (a) shows geometry and notations.
}
\label{fig:cone-spectrum}
\end{figure}
%==================================================================/

Let us analyse now the spin waves on the background of the axial state $\phi^{\mathrm{ax}} = \pm\pi/2$.
Similar to \eqref{eq:Lame-Bessel} one can find that
\begin{equation} \label{eq:linear-sep2}
\phi (\xi _1,\xi _2,t) \approx \pm\frac{\pi }{2} + e^{i\omega t + i\mu \xi _1 }\mathrm{P}(\rho ), \qquad \mu \in \mathbb{Z},
\end{equation}
where the radial function $\mathrm{P}(\rho ) = \mathcal{C}_1 \mathrm{J}_\nu (\mathfrak{q}\rho ) + \mathcal{C}_2 \mathrm{N}_\nu  (\mathfrak{q}\rho )$, with $\nu =\sqrt{\sin^2\psi + \mu ^2}/\cos\psi $. The boundary conditions \eqref{eq:bc} lead to the equation $\mathrm{J}_\nu'(\mathfrak{q}) \mathrm{N}_\nu '(\mathfrak{q} \rho _0) = \mathrm{J}_\nu '(\rho _0) \mathrm{N}_\nu '(\mathfrak{q})$, which determines the eigenfrequencies. Its numerical solutions for the lowest mode $\mu =0$ are plotted in Fig.~\ref{fig:cone-spectrum} for the case $\psi >\psi _c$. As well as in the previous case, the lowest frequency becomes arbitrary small with the cone size increasing. Nevertheless it is not so for the cylinder surface where the lowest frequency is fixed and it is equal to $\omega _c$. The case of cylinder ($\psi =\pi /2$) should be considered separately starting from the Eq.~\eqref{eq:dynamic}, whose linear solution against the axial state has the form $\phi =\pm \pi /2 + \mathcal{C} e^{i(\omega t + \mu \xi _1 + \mathfrak{q}_{\|}\xi _2)}$ with $\mathfrak{q}_{\|}$ being the wave vector along cylinder axis. The corresponding dispersion relation reads $\Omega  = \sqrt{1+\mu ^2+R^2\mathfrak{q}_{\|}^2}$. Existence of a gap in spectrum of the cylindrical magnetic shell was already predicted theoretically \cite{Gonzalez10} and checked by numerical simulations \cite{Yan11a}.

\section{Summary}
\label{sec:discussion} %

To conclude, we develop the general approach to describe the magnetisation states in arbitrary curved magnetic wires and shells in the  vanishing thickness limit. The curvature induces effective magnetic anisotropy and effective Dzyaloshinskii-like interaction.   We obtain an equation of magnetisation dynamics and propose a general static solution for the limit case of strong anisotropy. In the latter case the curvature effect is reduced to an influence of effective curvature induced magnetic fields. We illustrate our approach by two examples: (i) we calculate possible ground states of ring wires and compute the magnon spectrum in this system, (ii) we study the magnon spectrum in the cone shell. In both cases the curvature is the source of different possible ground states. The curvature contribution to the magnon spectrum of these systems is mostly due to the curvature induced anisotropy.

\appendix

\section{Exchange interaction of the curved wire: another representation}
\label{sec:energy-1d-another} %

In this appendix we discuss another angular parametrisation for the magnetisation:
\begin{equation} \label{eq:theta-phi-1D-another}
\vec{m} = \cos\Theta \, {\vec{e}}_1 + \sin\Theta \cos\Phi \, {\vec{e}}_2 + \sin\Theta \sin\Phi \,{\vec{e}}_3,
\end{equation}
where $\Theta =\Theta (s)$ and $\Phi =\Phi (s)$ are the angles in the Frenet–-Serret frame of reference: the polar angle $\Theta $ describes the deviation of magnetisation from the tangential curve direction, while the azimuthal angle $\Phi $ corresponds to the deviation from the normal. Similar to \eqref{eq:E-1D-theta-phi}, one can rewrite the energy terms as follows:
\begin{equation} \label{eq:E-1D-theta-phi-another}
\eqalign{
\mathscr{E}_{\mathrm{ex}}^{0} &= \left(\vec{\nabla}\Theta \right)^2 + \sin^2\Theta \left(\vec{\nabla}\Phi \right)^2,\\
\mathscr{E}_{\mathrm{ex}}^{A} &=(\kappa \cos\Theta \sin\phi -\tau \sin\Theta )^2+\kappa ^2\cos^2\Phi \\
\mathscr{E}_{\mathrm{ex}}^{D} &= 2\Phi '\sin\Theta \left(\tau \sin\Theta -\kappa \cos\Theta \sin\Phi \right)+2\kappa \Theta '\cos\Phi .
}
\end{equation}
Finally, the exchange energy takes the form, cf.~\eqref{eq:E-1D-final}
\begin{equation} \label{eq:E-1D-final-another}
\eqalign{
\mathscr{E}_{ex} \!=\! \left(\Theta ' \!+\!  \kappa  \cos \Phi \right)^2\!\! + \left[\sin\Theta \left(\Phi '\! + \! \tau \right)- \kappa \cos \Theta  \sin \Phi  \right]^2\!\!\!.
}
\end{equation}

%
%----------------------------------------------------------------
%
%\bibliography{soliton}
%
\providecommand{\newblock}{}

%

%----------------------------------------------------------------
%

\end{document}